\documentclass[twocolumn,showpacs,preprintnumbers,amsmath,amssymb,floats]{revtex4}
\usepackage{graphicx}
\usepackage{dcolumn}
\usepackage{bm}

\def\Re{\textrm{Re}}
\def\Im{\textrm{Im}}

\def\be{\begin{equation}}
\def\ee{\end{equation}}
\def\bea{\begin{eqnarray}}
\def\eea{\end{eqnarray}}

\begin{document}
\title{Topological Transition in the Fermi Surface of Cuprate
    Superconductors in the Pseudogap Regime}
\author{C. M. Varma and Lijun Zhu}
\affiliation{Department of Physics and Astronomy, University of
California, Riverside, California 92521, USA}

\begin{abstract}
The angle-resolved photoemission spectroscopy studies on cuprates in
the pseudogap region reveal an extraordinary topological transition
in which the ground state changes from one with a usual Fermi
surface to one with four Fermi points. Such a state is not possible
without some symmetry breaking which allows interference between
one-particle basis states which is normally forbidden. We also show
that the experimental results are quantitatively given without any
free parameters by a theory and discuss the implications of the
results.
\end{abstract}
\pacs{74.25.Jb, 74.20.-z, 74.72.-h}%
\maketitle

Recently, the single-particle spectral function $ A({\bf k},\omega)$
obtained by angle-resolved photoemission spectroscopy (ARPES) on
eight different underdoped BSCCO cuprates with $T_c$ ranging from
25K to 90K has been systematically analyzed \cite{kanigel}. The
principal conclusion is that the angular region in $\hat{{\bf k}}$
where $ A({\bf k},\omega)$ has a maxima at the chemical potential
$\mu$ is a universal function $\Phi(T/T_g(x))$. $T$ is the
temperature and $T_g(x)$ is the pseudogap transition temperature at
a given doping level $x$, the deviation of the density of holes from
half-filling. At $T\geq T_g(x)$, $\Phi=2\pi$, i.e. the angular
region encloses an area, while for $T \to 0$, $\Phi \to 0$ and the
region shrinks to 4 points. $T_g(x)$ is, within the uncertainties of
its determination, the same as that obtained from the resistivity or
the thermodynamic measurements such as the magnetic susceptibility
and the specific heat which also scale as functions of $T/T_g(x)$
\cite{tallon}. $T_g(x)$ is also consistent with the temperature at
which time-reversal symmetry (TRS) breaking is observed by dichroic
ARPES experiments \cite{kaminski, simon-varma} in the same
compounds. Qualitatively similar conclusions were arrived at by
Yoshida {\it et al.}\cite{fujimori}.

Recall that in a Fermi liquid \cite{pines-nozieres} [or a marginal
Fermi liquid (MFL) \cite{mfl}], $A({\bf k},\omega)$ is maximum at
$\omega=\mu$ for ${\bf k}={\bf k}_F$, thus defining the Fermi
surface.  The concept of a Fermi surface, properly defined as the
nonanalytic surface which separates the area of occupied states from
the unoccupied states in a Fermi liquid, is strictly meaningful only
at $T\to 0$. This is merely a technicality in the usual metallic
states. But for underdoped cuprates, the proper definition of a
Fermi surface is essential. The experimental results suggest that
for $x > x_c$, such that $T_g(x_c) = 0$, the extrapolated $T=0$
non-superconducting ground state has a Fermi surface, whereas for
$x<x_c$, the concept of a Fermi surface is lost.

It has become the custom to call the angular region $\Phi (T/T_g)$,
a {\it Fermi arc}. This is only harmless if the empirical procedure
used to derive  $\Phi (T/T_g)$ from the experiments is kept in mind.

The truly new physics is the deduction that at $T \to 0$, a gap in
the single-particle excitation spectra develops at the chemical
potential for a range of electronic density at all angles on the
Fermi surface except four nodal points. Of course, this deduction is
based on extrapolating data (available from about $T=400K$ to $25K$)
down to $T=0$, but the data (reproduced in Fig. \ref{fig:arc} below)
is persuasive. It joins the list of novel concepts brought to
physics by the cuprates because it does not obey Bloch's counting
theorem for periodic systems that gaps can only occur at Brillouin
zone boundaries, i.e., for integer fillings \cite{bloch, footnote1}.
The theorem originally derived for non-interacting electrons also
holds for interacting systems in which there is a one-to-one
correspondence between particles and quasiparticles. But the
derivation implies an even greater generality, because it depends
only on the (usually safe) assumption  that the excitation energy of
one-particle states is a single-valued continuous function of their
momentum quantum number except at the zone-boundaries, where
multiple energies are obtained for a given momentum due to
interference.

The only previously known state \cite{footnote2}, in the absence of
disorder, where a gap (or a gap with nodal points or lines as in
anisotropic superconductors) occurs tied to the chemical potential
independent of band-filling is superconductivity. In that case, the
energy of states is a double valued function of momentum for states
near ${\bf k}_F$ with a gap. Bloch's theorem is circumvented in this
case only because the single-particle excitations near ${\bf k}={\bf
k}_F$ are linear combinations of electrons and holes and therefore
are not eigenstates of charge. A gap occurs due to an interference
between basis states of different charges. Circumventing Bloch's
counting theorem in other circumstances requires that some other
normally sacrosanct quantum number be no longer protected, which
allows interference between basis states that is usually forbidden
\cite{footnote3}.

Bloch's theorem is circumvented in a theory of the pseudogap state
\cite{cmv99, cmv06} for reasons not having to do with BCS pairing.
This theory  predicts that the pseudogap state breaks time-reversal
symmetry below $T_g(x)$ without changing the translational symmetry.
This aspect has been tested in BSCCO by dichroic ARPES
\cite{kaminski, simon-varma} and in YBCO by polarized neutron
scattering \cite{fauque}. The statistical mechanical model derived
to obtain such a symmetry breaking \cite{aji-cmv} is the
Ashkin-Teller model, which in the relevant range of parameters has a
smooth change in specific heat at the transition with the entropy
released over a temperature range typically more than twice the
transition temperature \cite{sudbo}. It is also shown that a normal
Fermi surface cannot exist in such a TRS breaking state.  The order
parameter is not a conserved quantity; therefore, fermions with
crystal momentum ${\bf k}$ have a finite coupling $g({\bf k})$ to
the order parameter fluctuations in the long wavelength limit. Since
in this limit the energy of the order parameter fluctuations goes to
$0$, a Fermi surface instability with no change in translational
symmetry but in the harmonic of $g({\bf k})$ accompanies the broken
TRS. A stable state is found with an anisotropic gap in the
excitation spectra. In the new state, the single-particle
excitations are not eigenstates of crystal momentum but formed from
the linear combination of states of momentum in a small region
around a given momentum \cite{cmv99}. Interference between basis
states of crystal momentum  leads to the anisotropic gap at the
erstwhile Fermi surface with four Fermi points left intact. The
ground state itself is perfectly periodic and therefore an
eigenstate of crystal momentum. This is not new conceptually; recall
that in superconductivity the ground state conserves charge although
the single-particle excitations do not.

The purpose of this Letter is to show that the predictions of such a
state agree quantitatively with the new experimental results.
We
calculate the function ${\Phi}(T/T_g(x))$, defined above and compare
it with the experiments \cite{kanigel}. An independent experimental
result consistent with the underlying ideas is that the linewidth of
the single-particle spectra of the cuprates abruptly acquires a
large elastic part in going from the overdoped cuprates
\cite{kaminski-pr} to the underdoped cuprates.

The single-particle states at $T \to 0$ in the stable TRS breaking
state in the absence of impurity scattering have been derived
\cite{cmv99, cmv06} to have energies given by
\be
E^{\gtrless}_{\bf k} = \epsilon_{\bf k} \pm D({\bf k}),
\,\textrm{for}\; E_{\bf k} \gtrless \mu,
\label{eq:spec}
\ee
where $D({\bf k}) \approx
D_0(1-T/T_g)^{1/2}\cos^2(2\phi)/[1+(\epsilon_{\bf k}/\epsilon_c)^2]$
is the gap function, which is anisotropic and temperature-dependent.
$\phi$ is the angle of $\hat{\bf k}$ and $\epsilon_{\bf k}$ is the
``band-structure" energy of a tight-binding model  to fit the Fermi
surface \cite{band-structure} found by ARPES for $x\leq x_c$, with
effective nearest-neighbor and next-nearest-neighbor Cu-Cu hopping
parameters $t$ and $t'$ respectively:
\be
\epsilon_{\bf k} = - 2t \left[ \cos k_x + \cos k_y \right] -4 t'
\cos k_x \cos k_y,
\ee
where $t'/t=-0.35$. $\epsilon_c$ is a cutoff of order $D_0$. The
simple theory \cite{cmv06} gives $D_0 \approx \sqrt{6} T_g$.

The single-particle spectrum is given by the spectral function
\be
 A^{\gtrless}({\bf k}, \omega)= -{1\over\pi} {\Im \Sigma({\bf k},
\omega) \over [\omega-E^{\gtrless}_{\bf k}-\Re\Sigma({\bf k},
\omega)]^2 + [\Im\Sigma({\bf k}, \omega)]^2},
\label{eq:spectral}
\ee
where $\Sigma({\bf k},\omega)$ is the self-energy, which is the sum
of the contributions due to electron-electron scattering and
impurity scattering. We present here results in the pure limit as
well as including small angle impurity scattering. If $-\Im\Sigma
\gtrsim D(\hat{{\bf k}}_F)$, the spectrum $A^{>}({\bf k}, \omega)$
has finite weight in the $\omega<\mu$ region. So the total spectrum
measured by ARPES (at $\omega \leq \mu$) is
\be
A({\bf k}, \omega\leq\mu)=A^{<}({\bf k}, \omega)+A^{>}({\bf k},
\omega).
\label{eq:spec-def}
\ee

$\Sigma({\bf k},\omega)$  close to the maximum of $D(\hat{{\bf
k}})$, or far away from the {\it nodal} region can be calculated in
the pure limit from the fact that due to momentum and energy
conservation, decay of a particle requires three intermediate-state
particles, at least one of which must also be close to the maximum
of the gap function. Using these kinematical constraints and that
for $\omega, T \gtrsim D(\hat{{\bf
k}})$, the decay rate must revert to the state
without the pseudogap (i.e., the MFL state). So
\be
-\Im \Sigma_{in}({\bf k}, \omega, T) \approx \text{sech} \left( D(\hat{\bf
k}) \over \sqrt{\omega^2 + \pi^2 T^2} \right) \tau_M^{-1}(\omega,T)
\label{eq:sigma-in}
\ee
where $\tau_M^{-1}$ is the MFL relaxation rate:
\be
\tau_M^{-1} = \lambda \sqrt{\omega^2 + \pi^2
T^2}.
\label{eq:tau_mfl}
\ee
$\lambda$ is the dimensionless coupling constant used to fit ARPES
data for $x\geq x_c$ by the MFL spectral function.

Near the zeroes of $D(\hat{{\bf k}})$ (nodal region), one can again
use momentum and energy conservation to calculate the phase space
for decay. One easily finds that the self-energy for $\omega \ll
D(\hat{\bf k})$ is
\be
-\Im \Sigma_{in}({\bf k}, \omega, T) = \lambda {(\omega^2+\pi^2T^2)
/ D_0},
\label{eq:sigma-node}
\ee
while for $\omega \gg D_0$, it reverts to the MFL form
$\tau_M^{-1}$. Another interpolation form, such as in
Eq.(\ref{eq:sigma-in}) is used to connect
the nodal region and the antinodal region. In our calculations
below, we have used Eq.(\ref{eq:sigma-node}) in an angular region
extending to $\delta \phi=\pi/20$ from the nodal point and
Eq.(\ref{eq:sigma-in}) elsewhere. Similar results are obtained for
factors of 2 variations about this partitioning. For the low energy
region of interest, $\Re\Sigma$ produces negligible corrections and
has been ignored.

The spectral function is calculated using
Eqs.(\ref{eq:spectral},\ref{eq:spec-def}). The parameters used in
the evaluation are $D_0/T_g=2.5$, and $\lambda$ is fixed by the
value determined by the MFL fits to the spectral function for $x\geq
x_c$ to be $\approx 0.5$. Earlier, the specific heat and the
magnetic susceptibility in the underdoped cuprates were fitted
\cite{cmv06} to experiments with $D_0/T_g \approx 2.5$. Thus there
are no free parameters left to fit.

\begin{figure}[tbh]
\centering
\includegraphics[width=1.0\columnwidth]{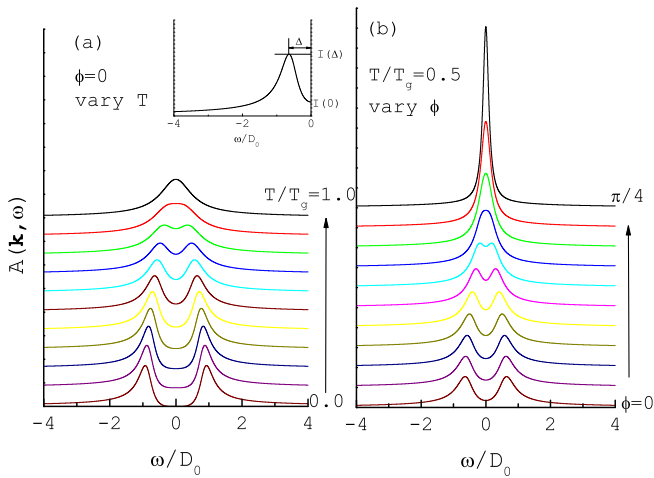}
\caption{The spectral functions (EDCs) for (a) the antinodal angle
at various temperatures; (b) fixed temperature ($T/T_g=0.5$) while
varying angles from antinode to node. Here, $D_0=2.5T_g$. As shown
in inset, we have followed the representation of the data as in
experiments, explained in the main text.}
\label{fig:spect}
\end{figure}

In Fig. \ref{fig:spect}, the calculated spectral function is plotted
at the anti-nodal point for various $T/T_g$ and for various angles
at $T/T_g = 0.5$. The inset shows the definition of various
quantities used to represent the experiments in Ref.\cite{kanigel}.
$\Delta(\phi)$ is defined as the
energy at which the spectral function peaks below the chemical
potential at the angle $\phi$, while $I(0,\phi)$ and
$I(\Delta,\phi)$ are the intensities at the chemical potential (set
to $\omega=0$) and at $\Delta$, respectively. Either
$\Delta(\phi)=0$ or $1-I(0,\phi)/I(\Delta,\phi)=0$ implies that
$A({\bf k}_F, \omega)$ peaks at the chemical potential, giving the
impression of "Fermi surface".
The angular regions where these quantities vanish are defined as
``Fermi arcs''.

\begin{figure}[tbh]
\centering
\includegraphics[width=0.9\columnwidth]{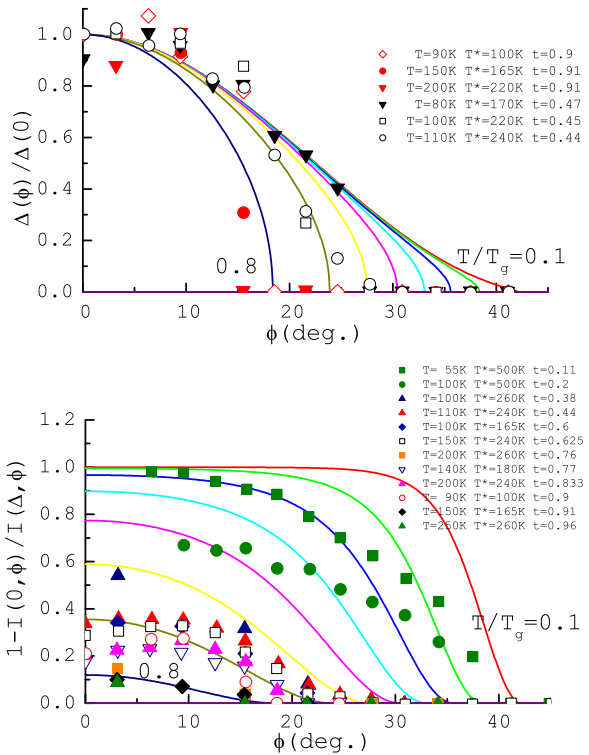}
\caption{The gap function and the peak intensity $1-I(0,\phi)/I(\Delta,\phi)$ as functions of the angle.
The solid lines are theoretical results for $T/T_g=0.1$ to $0.8$ (in
sequence from right to left).
} \label{fig:delta}
\end{figure}

The experimental results and the calculated values of
$\Delta(\phi)/\Delta(0)$ and of $1-I(0,\phi)/I(\Delta,\phi)$ are
shown in Fig. \ref{fig:delta}. The central aspect of the
experimental results deduced from the experiments, is shown in Fig.
\ref{fig:arc} together with the results of calculations. Fig.
\ref{fig:arc} shows that the Fermi arc length $\Phi(T/T_g(x))$ for a
whole range of underdoped cuprates at various dopings $x$ is a
universal function of $T/T_g(x)$ and that at $T \to 0, \Phi \to 0$.
The theory in the pure limit is in quantitative accord with the
experimental curve without any free parameters except for the region
of $T/T_g \approx 1$. This disagreement may be traced to the fact
that the experimental gap rises below $T_g$ much faster than the
mean-field theory. The qualitatively new physics in the results
deduced from the experiments is however in the low temperature
limit.

\begin{figure}[tbh]
\centering
\includegraphics[width=0.8\columnwidth]{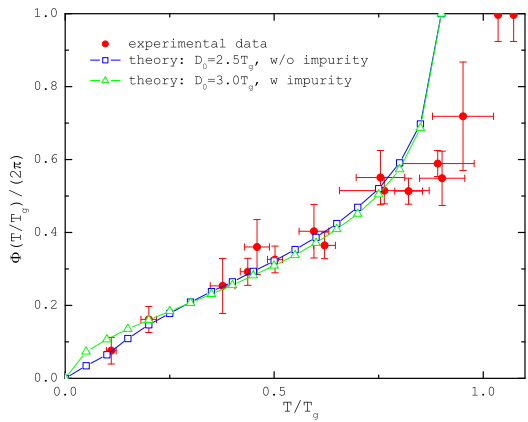}
\caption{The Fermi arc length as a function of $T/T_g$. The
experimental data from Ref.\cite{kanigel} are shown in red dots with
error bars. The solid lines are theoretical results: one in the pure
limit with $D_0=2.5T_g$; another with small angle impurity
scattering, fitted with $D_0=3 T_g$ and $1/\tau_0 = T_g$, where
$1/\tau_0$ is the impurity scattering rate at the nodal point, and
its magnitude can be estimated from the normal state data, such as
in Ref. \cite{kaminski-pr}.} \label{fig:arc}
\end{figure}

The agreement actually is a little worse if we include impurity
scattering, see also Fig. \ref{fig:arc}. We have calculated the
spectral function including the effects of small angle scattering
due to impurities between the Cu-O planes \cite{abrahams-varma-pnas,
abrahams-varma-pr}. Due to the anisotropy of the density of states
in the pseudogap state, the impurity scattering rate becomes
anisotropic as well as frequency-dependent. With the impurity
scattering estimated from experiments which gives an elastic
scattering rate of about $200 K$ at the nodal point above
$T_g$\cite{kaminski-pr}, we have to use $D_0/T_g \approx 3$ to get
agreement with the experiments. With small angle scattering, only
four Fermi points remain at $T\to 0$ but a weak smooth bump develops
for these values at around $T/T_g \approx 0.1$, which however
continues to give a theoretical curve within the experimental error
bars. The impurity scattering, of course, varies also from sample to
sample. We should also mention that if the large angle impurity
scattering in the unitary limit were important, say due to
impurities in the plane, we expect an impurity resonance at the
nodal points, so that the Fermi points are smeared out.

We are not aware of any other theory giving the results of the
experiments, specifically a scaling form for the Fermi arc
$\Phi(T/T_g(x))$ with $\Phi(0) \to 0$.  There exist theories which
generate anisotropic reduction of the density of states in the
underdoped state without generating Fermi points at $T=0$, such as
approximate solutions of the Hubbard model with dynamical mean-field
theory and its extensions \cite{tremblay}. There is no indication
that the Fermi surface shrinks to four Fermi points in such
calculations at $T \to 0$. This is consistent with the general
arguments given in this paper that this is not possible without some
symmetry breaking. On the other hand, experiments must be done in
materials where the extrapolation to the ground state of the
pseudogap phase can be done more precisely. This necessitates
experiments in high quality underdoped samples just at the boundary
to superconductivity. Another prediction of the theory which awaits
more precise measurements through higher energy and momentum
resolution is that the angular dependence of the pseudogap is
$\propto (\phi-\phi_0)^2$ near the points $\phi_0$ at which it is
$0$, [see the expression for the pseudogap following Eq.(1)]. This
is important to conclude that there is no phase associated with the
pseudogap, unlike superconducting gaps or gaps due to change in
translational symmetry due to charge or spin-density waves or
staggered flux phases.

In conclusion, we would like to re-emphasize that the scaling of the
Fermi arc is a new phenomena in condensed matter physics and that it
requires a broken symmetry. We have presented the results of a
theory, one of whose consequences is a Fermi surface with only four
points. With two parameters taken to fit thermodynamic measurements,
we have been able to account for the details of the Fermi arcs. The
same theory has predicted a time-reversal violating state in the
pseudogap region and the theory of quantum critical fluctuations
\cite{aji-cmv} which gives the phenomenological spectrum \cite{mfl}
with which the ``strange metal'' phase is understood. These
fluctuations also give rise to attractive paring interaction in the
``d-wave'' symmetry \cite{super}.

{\it Acknowledgement}. One of us (CMV) is grateful to Leon Balents
for a discussion of the Bloch's counting theorem.


\begin{thebibliography}{99}
\bibitem{footnote1}
The theorem continues to hold for gaps which are zero on points and
lines because when integrating over the Brillouin zone, points or
lines contribute negligibly to the number of total states.
\bibitem{footnote3}
A change of the connectivity of the Fermi surface without change of
symmetry was described by Lifshitz \cite{lifshitz, volovik} many
years ago. Clearly, the shrinking of the Fermi surface to four
points, while retaining all the symmetries is a topological
transition. But in a Lifshitz transition, the total volume of the
Fermi surface (taking into account the ``sign" of the Fermi surface,
positive for electrons and negative for holes) is conserved. The
(extrapolated) ground state in the underdoped cuprates encloses no
electrons or holes.
\end{thebibliography}
\end{document}